\documentclass[sn-mathphys,Numbered]{sn-jnl}% Math and Physical Sciences Reference Style
\usepackage{graphicx}%
\usepackage{multirow}%
\usepackage{amsmath,amssymb,amsfonts}%
\usepackage{amsthm}%
\usepackage{mathrsfs}%
\usepackage[title]{appendix}%
\usepackage{xcolor}%
\usepackage{textcomp}%
\usepackage{manyfoot}%
\usepackage{booktabs}%
\usepackage[ruled,vlined]{algorithm2e}%
\usepackage{algorithmicx}%
\usepackage{algpseudocode}%
\usepackage{listings}%
\usepackage{siunitx}%
\usepackage{subcaption}
\begin{document}

\title[4D Track Reconstruction on Free-Streaming Data at PANDA at FAIR]{4D Track Reconstruction on Free-Streaming Data at PANDA at FAIR}


\author*[1,2]{\fnm{Jenny} \sur{Taylor}}\email{j.taylor@gsi.de}

\author[1]{\fnm{Michael} \sur{Papenbrock}}

\author[3]{\fnm{Tobias} \sur{Stockmanns}}

\author[2,4]{\fnm{Ralf} \sur{Kliemt}}

\author[1]{\fnm{Tord} \sur{Johansson}}

\author[1]{\fnm{Adeel} \sur{Akram}}

\author[1]{\fnm{Karin} \sur{Schönning}}

\subtitle{On Behalf of the PANDA Collaboration}

\affil*[1]{\orgname{Uppsala University}, \orgaddress{\city{Uppsala}, \country{Sweden}}}

\affil*[2]{\orgname{GSI Helmholtzzentrum für Schwerionenforschung GmbH}, \orgaddress{\city{Darmstadt}, \country{Germany}}}

\affil[3]{\orgname{Forschungszentrum Jülich GmbH}, \orgaddress{\city{Jülich}, \country{Germany}}}

\affil[4]{\orgname{Ruhr-Universität Bochum, Institut für Experimentalphysik I}, \orgaddress{\city{Bochum}, \country{Germany}}}


\abstract{A new generation of experiments is being developed, where the challenge of separating rare signal processes from background at high intensities requires a change of trigger paradigm. At the future PANDA experiment at FAIR, hardware triggers will be abandoned and instead a purely software-based system will be used. This requires novel reconstruction methods with the ability to process data from many events simultaneously. 

A 4D tracking algorithm based on the cellular automaton has been developed which will utilize the timing information from detector signals. Simulation studies have been performed to test its performance on the foreseen free-streaming data from the PANDA detector. For this purpose, a quality assurance procedure for tracking on free-streaming data was implemented in the PANDA software. The studies show that at higher interaction rates, 4D tracking performs better than the 3D algorithm in terms of efficiency, 84$\%$ compared to 77$\%$. The fake track suppression is also greatly improved, compared to the 3D tracking with roughly a 50$\%$ decrease in the ghost rate. }

\keywords{4D Tracking, PANDA, Quality Assurance, Free-Streaming Data, Cellular Automaton}



\maketitle

\section{Introduction}
\label{sec:introduction}

The search for rare processes and previously unknown particles push the frontier of modern experimental hadron physics to higher and higher reaction intensities. We are even about to reach a point where our ability to produce abundant data outpaces our ability to retain them. This fact is correlated with the rapid increase in processing power of computing hardware, in contrast to the slower evolution of data transport and storage.
As a result of the ever-improving accelerators and detectors world-wide, the past decades have seen a remarkably rapid progress, with discoveries such as the Higgs boson \cite{higgs1,higgs2} and exotic multiquark states \cite{PhysRevLett.91.262001, PhysRevLett.95.142001, PhysRevLett.110.252001}. In order to solve current puzzles in hadron and particle physics, a new generation of facilities is underway. For all of these, a crucial component will be to discriminate between signals of interest and background at an early stage.

For a long time, hardware triggers served this purpose very well, selecting potentially interesting physics events by identifying specific signatures \textit{e.g.} high energy loss, and storing the corresponding events for reconstruction and more refined filtering later. As more and more of the low-hanging fruits of subatomic physics have been harvested, the hunting ground for unknown physics has to a large extent become processes whose signatures are expected to be very similar to those of background. Hence, more sophisticated data filtering methods are required -- and the hardware trigger paradigm falls short. 
The new generation of facilities is responding to this challenge by moving to different variations of a trigger-less data acquisition, having data streaming continuously from the detectors and selecting the data only after they have been reconstructed online during their acquisition.
These approaches are referred to as \textit{software triggers}.

 Currently operating and upcoming experiments are embracing the new paradigm in various degrees. The high-level trigger of the ALICE experiment performs the full event reconstruction, calibration and high-level data quality assurance in almost real time \cite{ALICE:2018phe}. The LHCb collaboration is currently making efforts to implement full online analysis chains in the software trigger apart from computing observables for the event filtering \cite{LHCb:2018zdd}. XENON1T has been collecting data altogether with a software-trigger \cite{XENON:2019bth}. At the future Facility for Antiproton and Ion Research (FAIR), the CBM \cite{Hohne2011} and PANDA \cite{PANDA:2009yku} experiments will also employ fully software-based trigger systems. %In this work, a new solution for tracking for this purpose has been implemented. 

The introduction of software triggers calls for a different treatment of the detector signals. In the traditional approach, data are processed event-by-event in the form of contained data packages, containing all detector signals collected within a pre-defined time-window as a response to the hardware trigger. In the approach needed for free-streaming data, a continuous stream of data is processed and organised based on spatial and timing information. Tracks can then be formed \textit{e.g.} by 4D track reconstruction where time is used in addition to spatial coordinates. A 4D Cellular Automaton (CA) is \textit{e.g.} utilized at CBM \cite{CBMCA} with excellent performance. 

Track reconstruction algorithms are evaluated using Quality Assurance (QA) methods. For tracking event-by-event, there are established methods such as those used and outlined in Ref. \cite{walter}. However, QA methods for tracking on free-streaming data are hard to find within hadron and HEP literature. 
%\textcolor{blue}{Quality assurance: cite some previous work, for example from Belle II and also Walter's paper, to give at least some kind of a state of the art. The conclusion should be that to our best knowledge, until now there has not been any QA method for 4D tracking. This strengthens the value of this paper since it becomes clearer that we add something new, of general interest.}

In this article, a 4D CA track finding algorithm developed for a continuous stream of hit data from the PANDA experiment is presented. In addition, a quality assurance method to evaluate the performance of the track finding was developed which is the first of its kind and is outlined in detail in this article.

\section{The PANDA experiment}
\label{sec:panda_experiment}

\begin{figure*}
    \centering
    \includegraphics[width=1\linewidth]{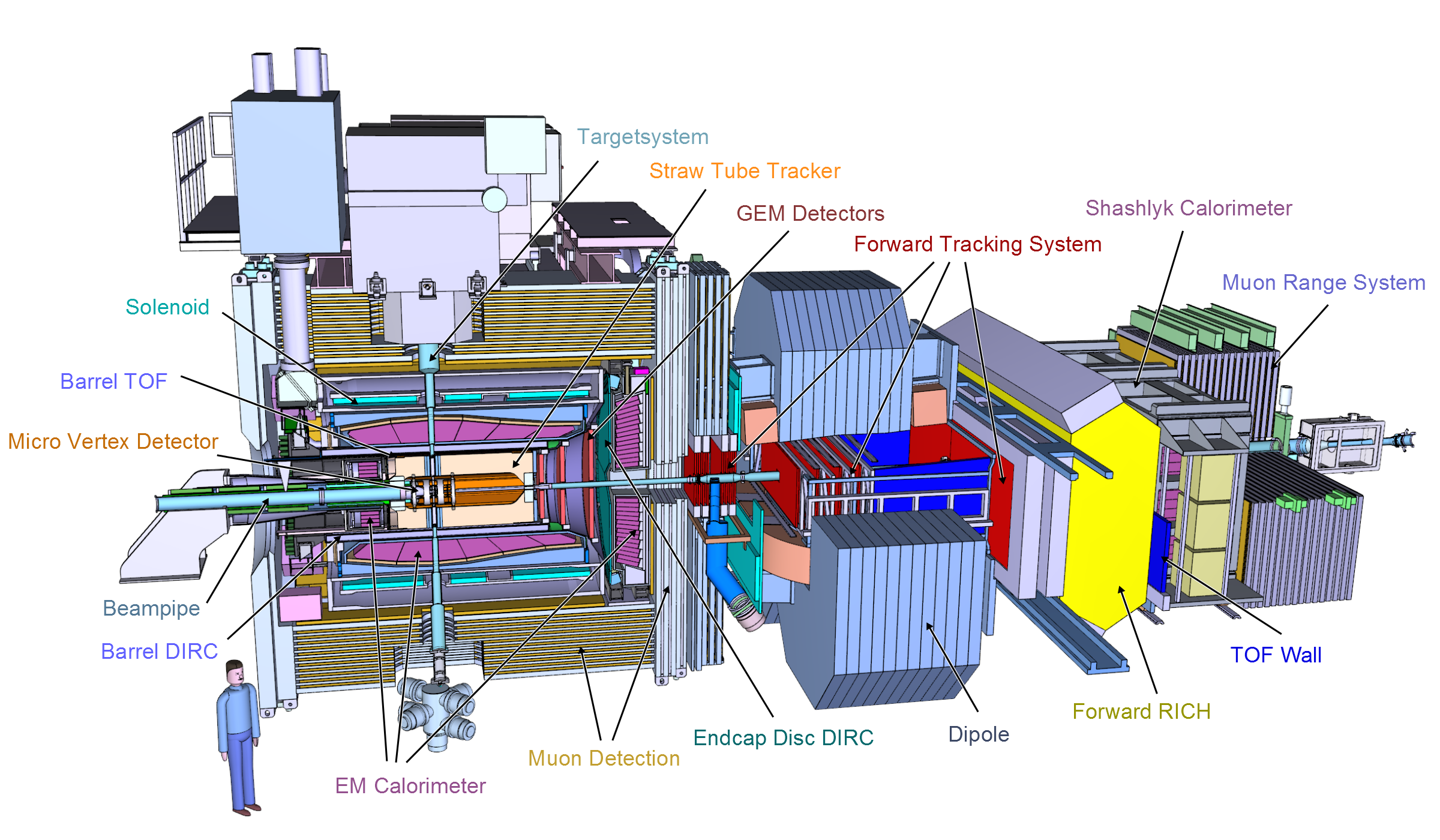}
    \caption{The PANDA detector. All subdetectors are labeled.}
    \label{fig:panda_detector}
\end{figure*}

PANDA (antiProton ANnihilations at DArmstadt) \cite{PANDA} is a future multipurpose detector designed to investigate QCD in the energy region where quarks form hadrons. The broad physics program includes four main pillars: strangeness physics, charm and exotics, nucleon structure and hadrons in nuclei \cite{phaseone}. 

The antiproton beam will be delivered by the High Energy Storage Ring (HESR), a racetrack-shaped ring with a circumference of \SI{575}{\metre}. The beam will cover a momentum range from 1.5 GeV/\textit{c} to 15.0 GeV/\textit{c}. In Fig.~\ref{fig:panda_detector}, the beam enters from the left.

The detector, shown in Fig.~\ref{fig:panda_detector}, is divided into a target solenoid spectrometer surrounding the interaction point and a forward dipole spectrometer to detect particles emitted at small scattering angles. Both spectrometers comprise subdetectors for tracking, particle identification and calorimetry.  The track reconstruction algorithm presented in this work treats signals from the central Straw Tube Tracker (STT) \cite{PANDA:2013jpu} and can be extended to also include the Micro Vertex Detector (MVD) \cite{PANDA:2012hbk}, the Gas Electron Multipliers (GEM) \cite{DivaniVeis:2018kak} and the Forward Tracking System (FTS) \cite{PANDA:2017wqo}. 
The coordinate system is defined with the \textit{z}-direction along the beam line and the \textit{xy}-plane perpendicular to the beam line. 

\subsection{The Straw Tube Tracker}
\label{sec:stt}

\begin{figure}
    \centering
    \includegraphics[width=1\linewidth]{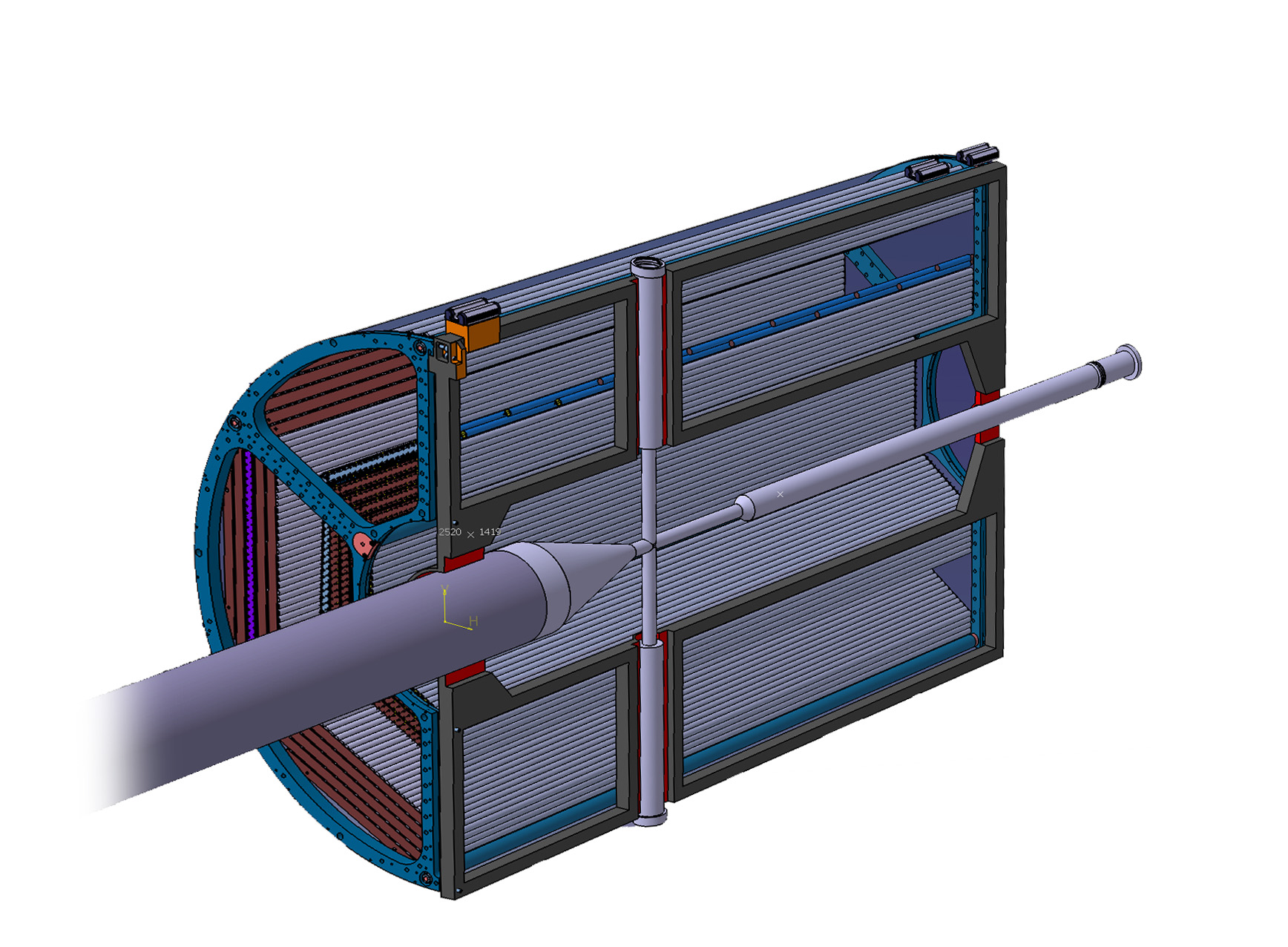}
    \caption{Cutaway view of the Straw Tube Tracker. The detector surrounds the beamline and has vertical cavities for the target beamline.}
    \label{fig:stt}
\end{figure}

The main tracking detector of the central part of PANDA is the STT, Fig.~\ref{fig:stt}. It consists of 4224 closely packed single-channel readout drift tubes placed in a hexagonal pattern. The detector has a cylindrical shape with roughly 27 tube layers in the radial direction, 19 parallel layers for \textit{xy}-reconstruction and 8 central layers of tilted tubes, $\pm$2.9$^{\circ}$ with respect to the beam line, for \textit{z}-reconstruction. The detector will be placed between \SI{15.0}{\centi\metre} and \SI{41.8}{\centi\metre} in radial direction and the tubes are \SI{140.0}{\centi\metre} long with a diameter of \SI{1.0}{\centi\metre}. The tubes will be filled with a 90/10 gas mixture of $\text{Ar/CO}_\text{2}$. The walls consist of a \SI{27}{\micro\metre} thick layer of aluminized mylar foil and the tubes will have \SI{20}{\micro\metre} thick anode wires in the center that are made of gold-plated tungsten.

When a charged particle traverses a tube, the gas will be ionized and free electrons are created along the trajectory of the particle. An applied electric field will make the electrons accelerate towards the anode wire. By measuring the arrival time of the signal precisely and in combination with other detectors it is possible to determine the \textit{drift} time of the electrons. Based on that the \textit{isochrone} radius is calculated, a circle centered at the tube wire and going through the point of closes approach of the particle to the wire. The maximum drift time, \textit{i.e.} the time it takes the electron to drift to the wire, is around \SI{250}{\nano\second} for electrons created close to the tube wall. 

%\subsection{Track reconstruction at PANDA}
%\label{sec:track_reconstruction_panda}

%Several track reconstruction algorithms have been developed for PANDA and many use the STT information as a starting point. \textcolor{red}{Review of some previous algorithms. In STT: SttCellTrack, Riemann track fitting, Triplet finder on GPU (Herten), tracking on FPGA (Yutie), Pz-finder paper, hough+apploonius. Mention Genfit2 and Kalman filter? In forward direction, mention Waleeds algorithm.}

\subsection{Free-streaming detector data}
\label{sec:free_streaming_data}

 The HESR antiproton beam will be quasi-continuous, and give rise to Poisson distributed instantaneous interaction rates when interacting with the fixed internal proton or nuclear target. The revolution time of the beam is around \SI{2 000}{\nano\second}, corresponding to a frequency of \SI{500}{\kilo\hertz}, but the exact number depends on the beam momentum. The beam will be active during 80$\%$ of the time and a gap with no beam during 20$\%$. This structure creates natural separation for data packaging.

During the first phase of operation, the luminosity will be 2 $\times$ 10$^{\text{31}}$ cm$^{\text{-1}}$s$^{\text{-1}}$ resulting in an average interaction rate of 2.0 MHz. The final design luminosity of PANDA is 2 $\times$ 10$^{\text{32}}$ cm$^{\text{-1}}$s$^{\text{-1}}$ resulting in an average interaction rate of \SI{20.0}{\mega\hertz}. The raw data rate before the software trigger will be around 200 GB/s. This needs to be reduced by a factor of 1000.

\begin{figure}
    \centering
    \includegraphics[width=1\linewidth]{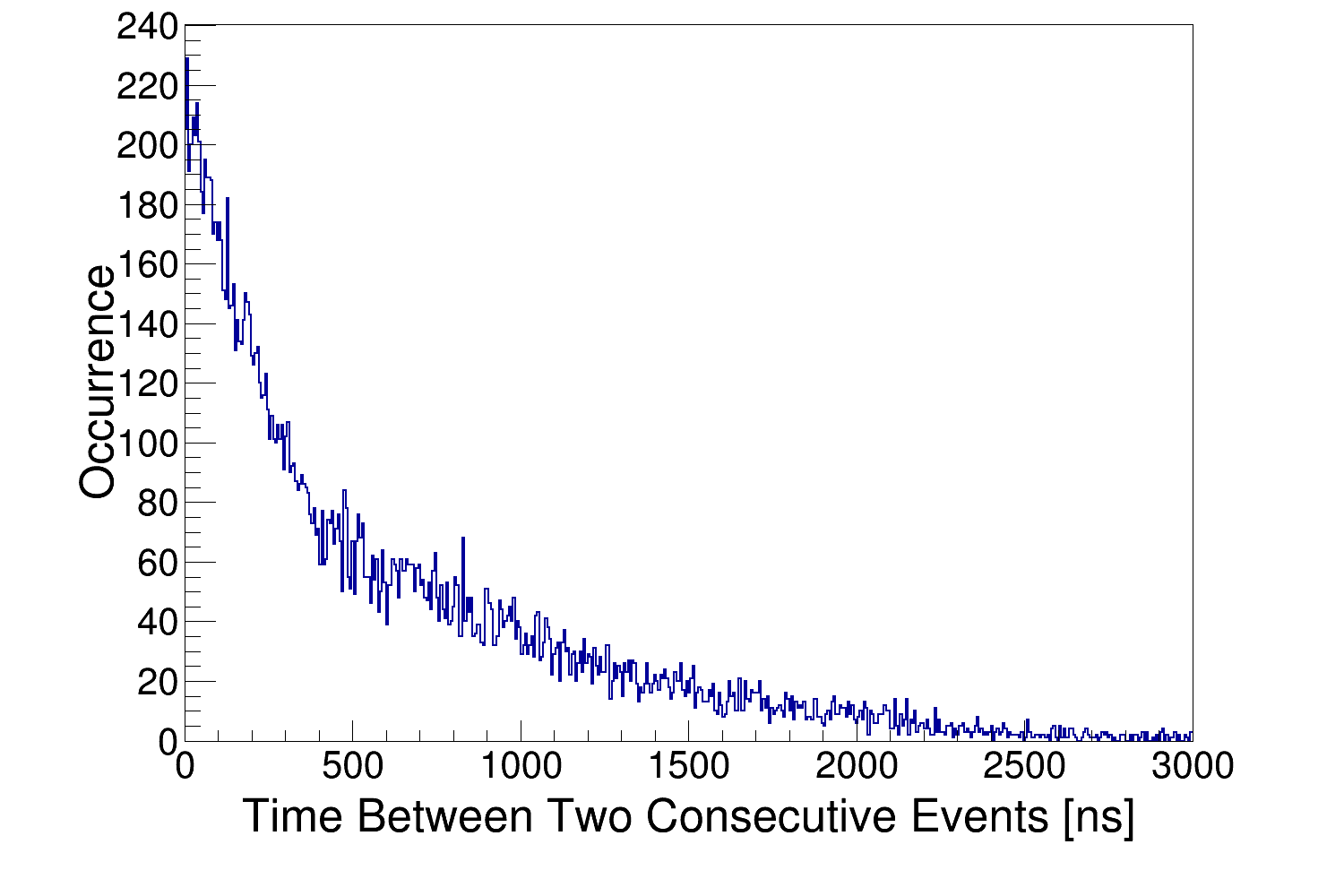}
    \caption{The time between the start of two consecutive events at 2.0 MHz. The average time is 500 ns at this interaction rate. Many events occur closer in time as well as more separated.}
    \label{fig:time_between_events}
\end{figure}

The HESR beam structure in combination with the foreseen high intensities implies that signals originating from different physics reactions will overlap in time. At an average interaction rate of \SI{2.0}{\mega\hertz}, the average time between the start of two consecutive events will be \SI{500}{\nano\second}. In Fig.~\ref{fig:time_between_events}, it is however shown that many consecutive events are separated by much less than \SI{500}{\nano\second}. At the PANDA design interaction rate of \SI{20.0}{\mega\hertz}, the average time between two events will be \SI{50}{\nano\second}. 

\begin{figure*}
    \centering
    \includegraphics[width=0.49\linewidth]{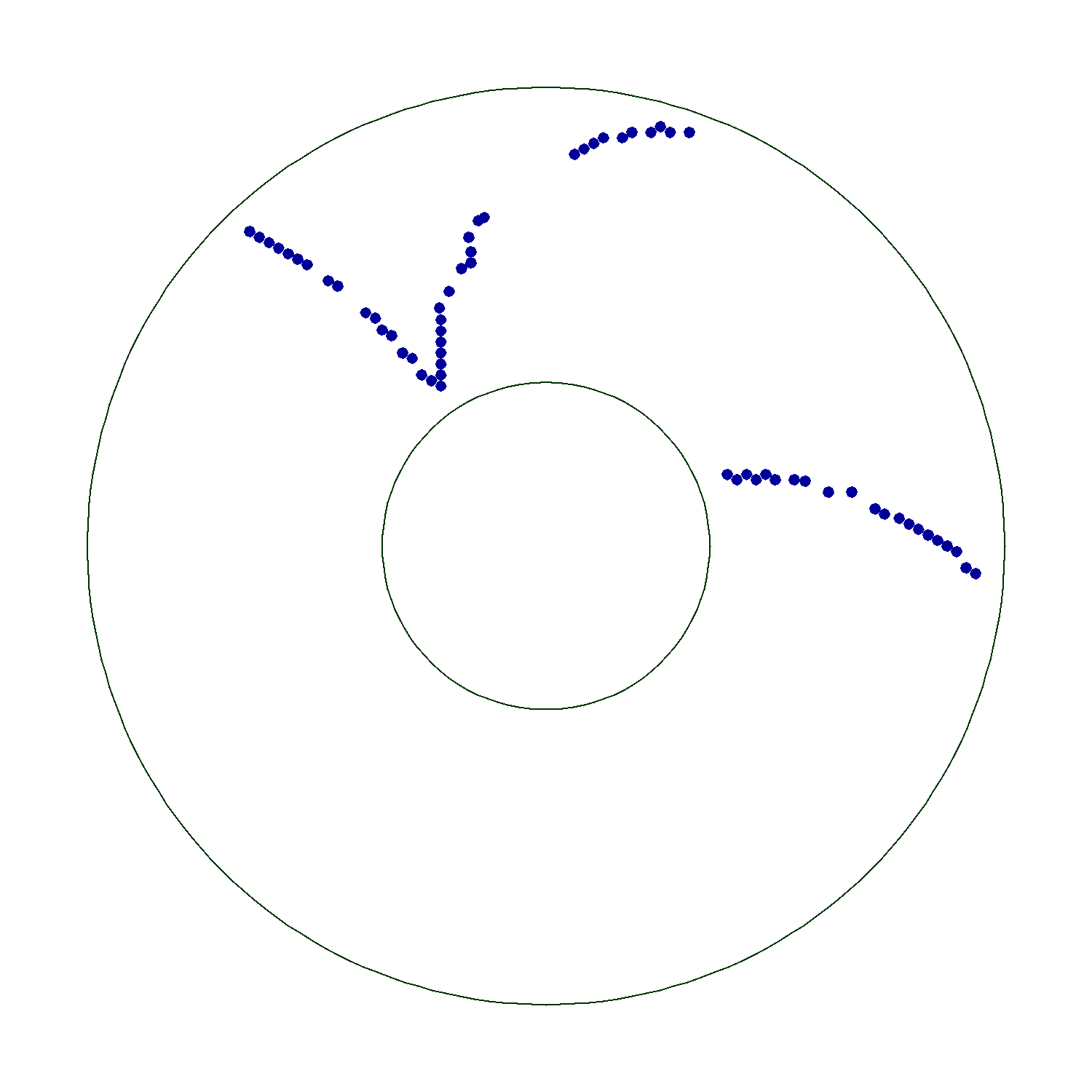}
    \includegraphics[width=0.49\linewidth]{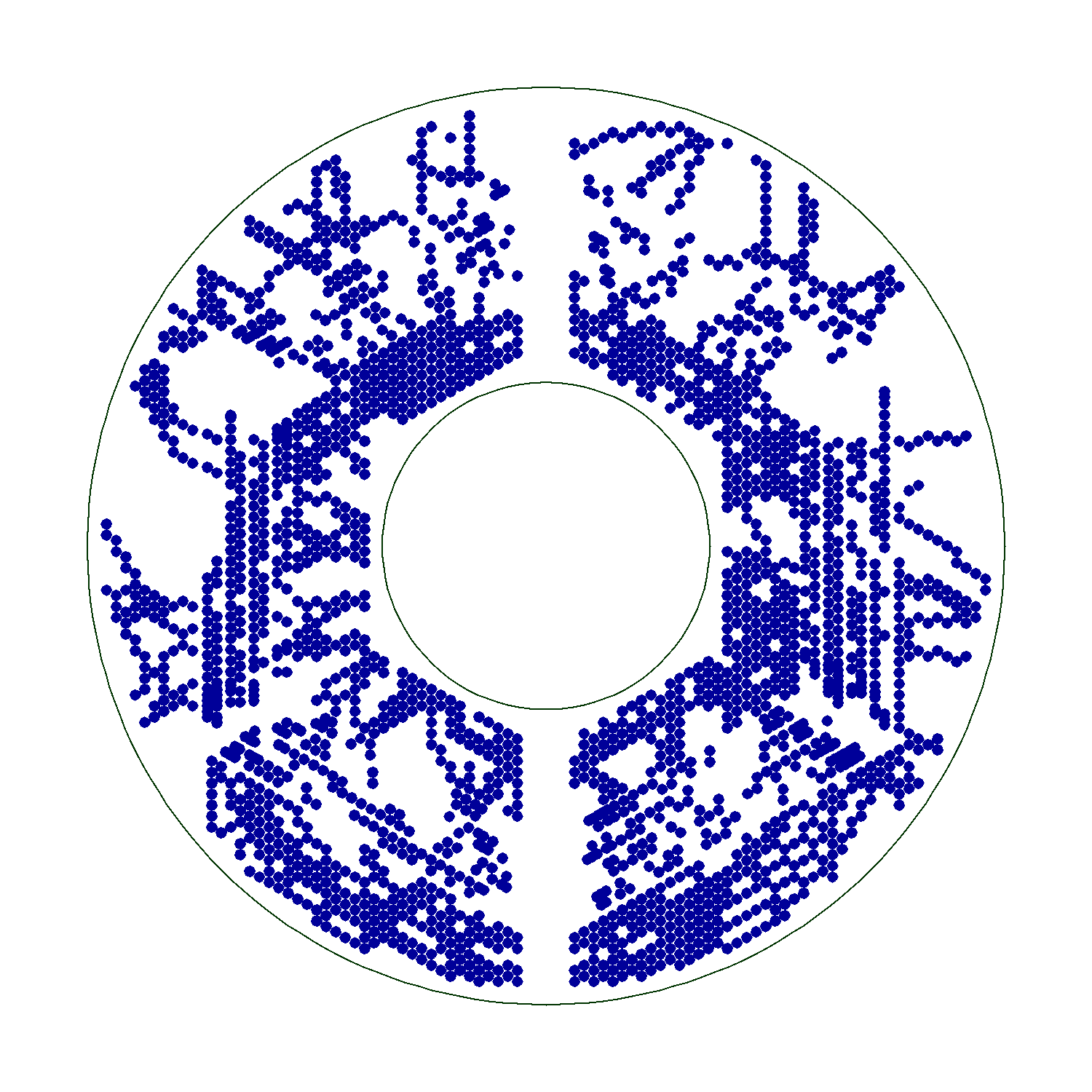}
    \caption{Detector occupancy of the PANDA Straw Tube Tracker at low intensities, where events do not overlap in time (left) and high intensities, where up to 40 events overlap (right).}
    \label{fig:low_high_intensity}
\end{figure*}

 Each subdetector at PANDA will be self-triggering so the detector information is read out independently of other subdetectors. The information is then synchronized and sorted in time by using the time-stamp each detector has to assign to its data. The reconstruction and event building can subsequently take place. This, combined with the Poisson distributed interaction rates, means that the hits from several events will be processed simultaneously. Since the STT has a dift-time of up to 250 ns, signals from different events will have a higher tendency to overlap in time in this detector compared to others, as illustrated in Fig.~\ref{fig:low_high_intensity}.

\subsubsection{Simulations for Free-Streaming Data}
\label{sec:FreeStreamingData}
\begin{figure*}
    \centering
    \includegraphics[width=1\linewidth]{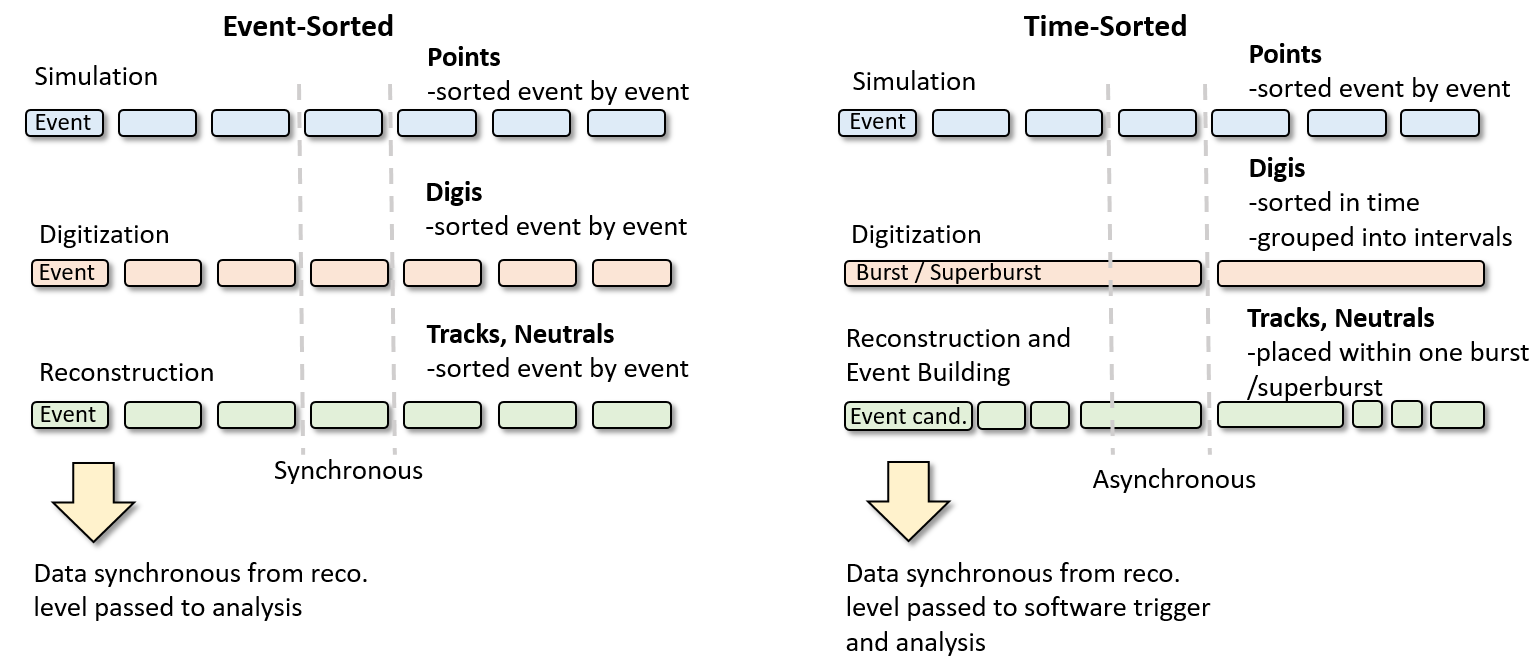}
    \caption{The event-sorted simulation (left) compared to the time-sorted simulation (right). The event-sorted simulation is synchronous until the software trigger and the time-sorted simulation is asynchronous until the reconstruction and synchronous following that. One burst of data corresponds to events within a certain time-interval and a superburst consists of a number of bursts}
    \label{fig:time_based_and_event_based_simulation}
\end{figure*}

The free-streaming data is simulated in FairRoot \cite{ContinuousReadout} upon which the PANDA software PandaRoot \cite{PandaRoot} builds. Fig.~\ref{fig:time_based_and_event_based_simulation} illustrates the difference between time-sorted and event-sorted simulations. In both cases, the events are generated one-by-one at the \textit{simulation stage}. This is followed by the \textit{digitization stage} where the output from the simulation is converted to signals that realistically mimics the detector output, including spatial resolution, detector granularity, timestamps and pulse shapes. At this stage, event-sorted and time-sorted simulations differ: while in the first case, the digitized hits are sorted event-wise, in the time-sorted case, the digitized hits are sorted according to time instead, and passed on to the next \textit{reconstruction} stage in pre-defined time intervals into \textit{bursts}. Hence, the event-sorted simulation chain is synchronous while the time-sorted chain is asynchronous. The latter means that data reconstruction operates on detector hits coming from several events simultaneously. 

\begin{figure*}
    \centering
    \includegraphics[width=0.49\linewidth]{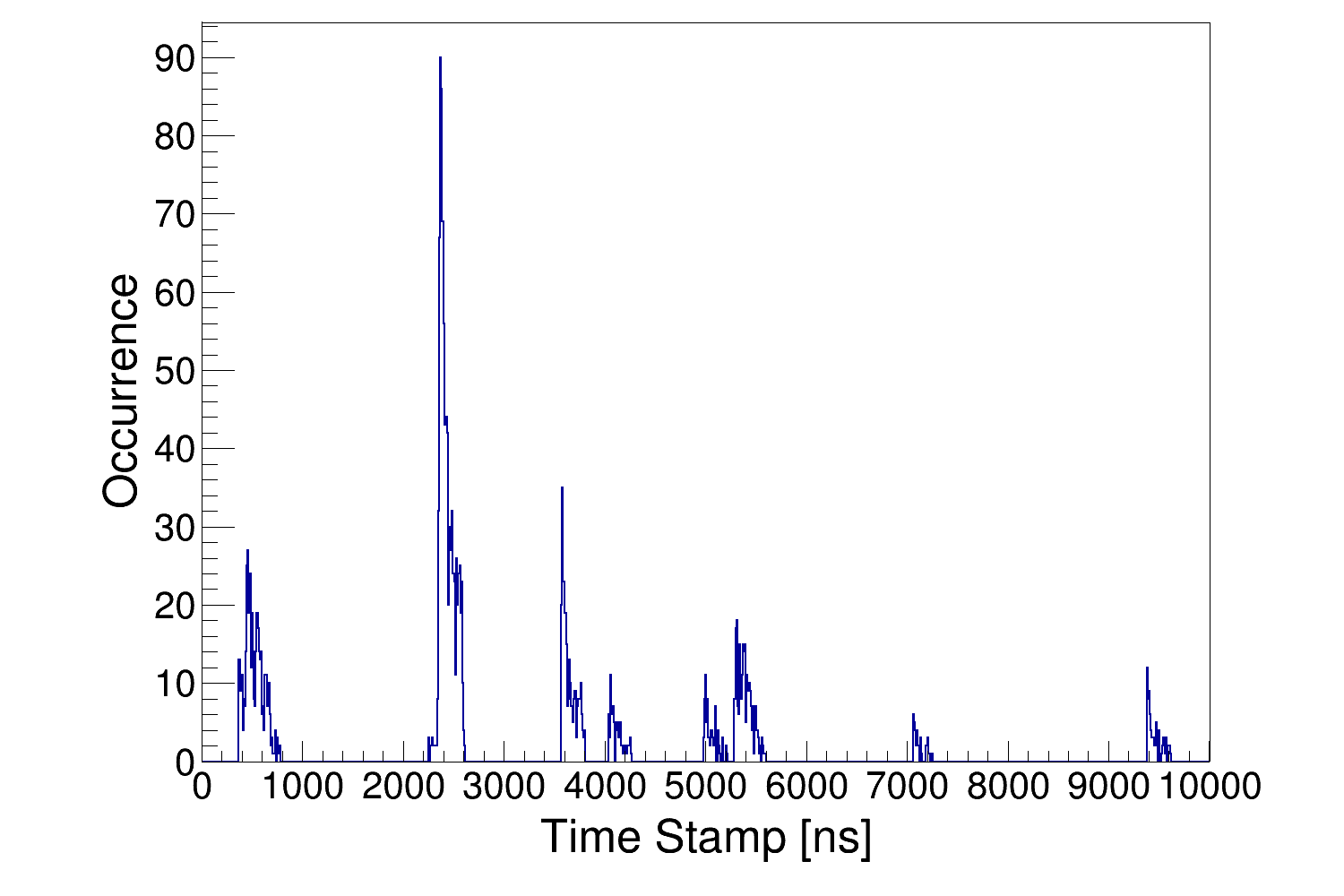}
    \includegraphics[width=0.49\linewidth]{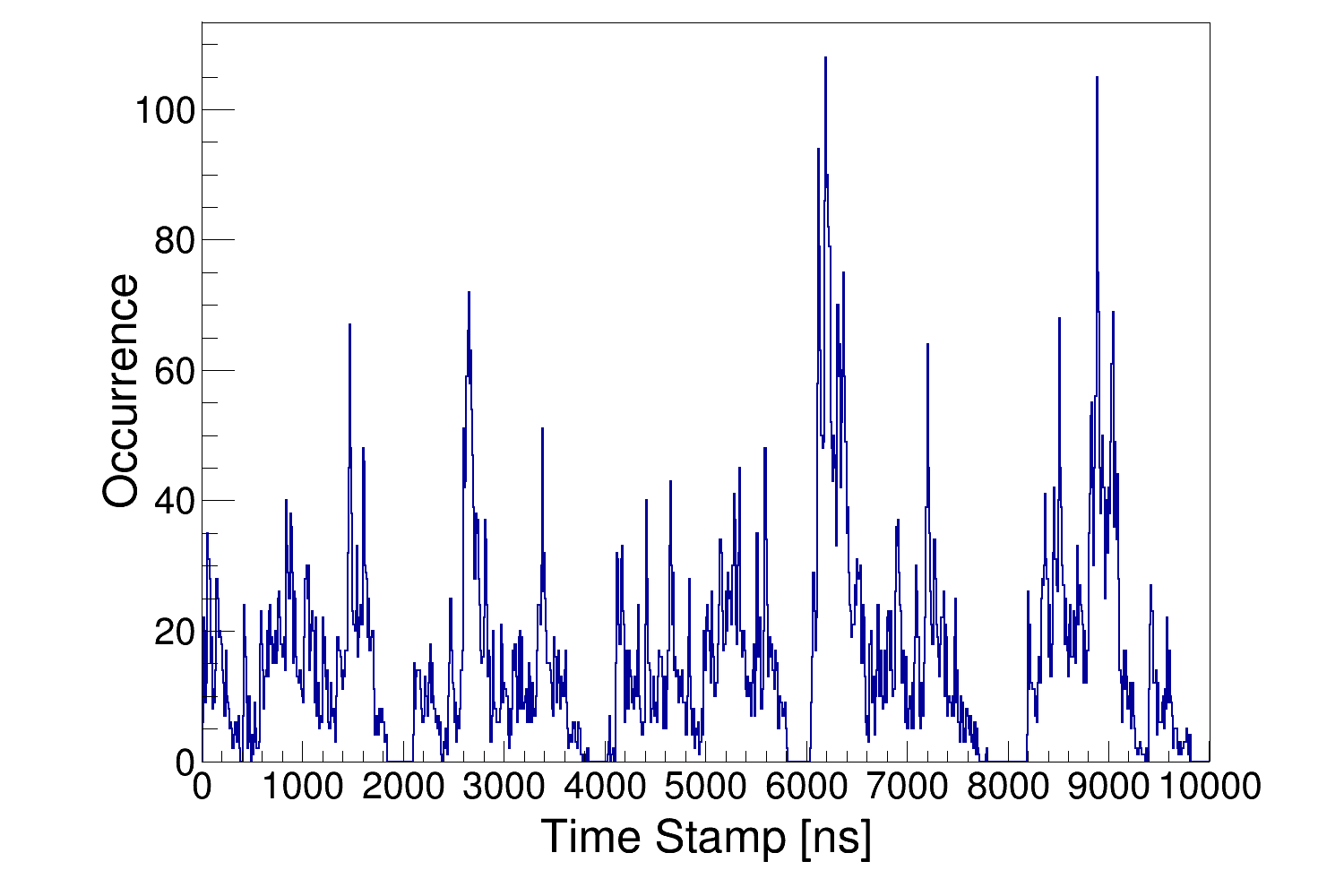}
    \caption{Time-stamps of the hits in the STT at 2.0 MHz (left) and 20.0 MHz (right). The time-stamp peaks at 2.0 MHz correspond roughly to one event.}
    \label{fig:event_structture_stt}
\end{figure*}

When processing time-sorted data, a time interval for obtaining the hits must be chosen. In the PANDA STT detector, the time structure of the hits is shown in Fig.~\ref{fig:event_structture_stt} for two cases. At an interaction rate of 2.0 MHz (left panel), the events are fairly well separated in time since the average time between two events exceeds the maximum drift time of the electrons. However, at 20.0 MHz (right panel), signals from different events are overlapping in time. The larger gaps at around 2000, 4000, 6000, 8000 and 10000 ns reflect the time structure of the beam. 

In the reconstruction of free-streaming data, the data can be grouped in certain time intervals. For this purpose, two methods are available in PandaRoot. The first method is to retrieve all data within a fixed time interval. This leads to a structure of bursts of data of the same length in time. The second method is to retrieve all data between two time gaps in the data stream have been found. This leads to bursts of varying length in time and assumes that the data of one event is close in time and that larger time gaps in the data stream can be used to separate events from each other. In an event-by-event simulation the data is retrieved already grouped into the correct event structure.

%The free-streaming data structure requires dedicated tracking algorithms and quality assurance methods, in particular when the former are applied as part of a software trigger scheme.

\section{4D track reconstruction}

The advent of fast silicon detectors has led to the introduction of 4D track reconstruction, where a timestamp is assigned to each point (or hit) along the track. In environments with high track multiplicity, timing information is expected to help distinguishing between hits from different particles.

In the following, an algorithm is presented that was originally developed to operate on event-sorted data but later updated for time-sorted data. This track finder combines a CA \cite{Wolfram:1982me} for pattern recognition with a Riemann paraboloid fitting  procedure~\cite{RiemannParaboloid} for track parameter estimation. The algorithm is based on pattern recognition in the PANDA STT, and is general in the sense that it does not require tracks to originate from the beam-target interaction point. Hence, it should be able to reconstruct tracks from secondary vertices, such as tracks from hyperon decay products. In this work, the algorithm was developed to perform tracking in 4D, utilizing the timestamp of each hit and to work on time-sorted data.
%It is named the \textit{SttCellTrackFinder} \textcolor{red}{Ref sttcelltrack?}.

\subsection{Track reconstruction event-by-event}
\label{sec:even_based_reconstruction}

%The time-based tracking algorithm builds on an event-based algorithm which utilizes a cellular automaton (CA)  to identify patterns. 

The CA-based tracking algorithms operate on a grid of cells, which can have \textit{e.g.} a 1D or a 2D/3D rectangular or hexagonal layout. The state of a cell is determined by the state of its neighbours. Since CA algorithms are inherently local, they are suitable for parallelization which has made them popular for track reconstruction purposes. 

Given the tightly packed, hexagonal layout of the STT, CA algorithms are suitable for track finding in this detector and was first applied in Refs \cite{sttcelltrack1,sttcelltrack2}. Here, each STT tube translates into a cell.

In CA algorithms, the state of a cell is either active or inactive. In this case with the STT tracker, active means that a hit is read out from the corresponding straw. Whether an active cell is a part of a pattern or not is uniquely determined by the states of its neighbours. 

\begin{figure*}
    \centering
    \begin{subfigure}{0.2\textwidth}
        \includegraphics[width=\linewidth]{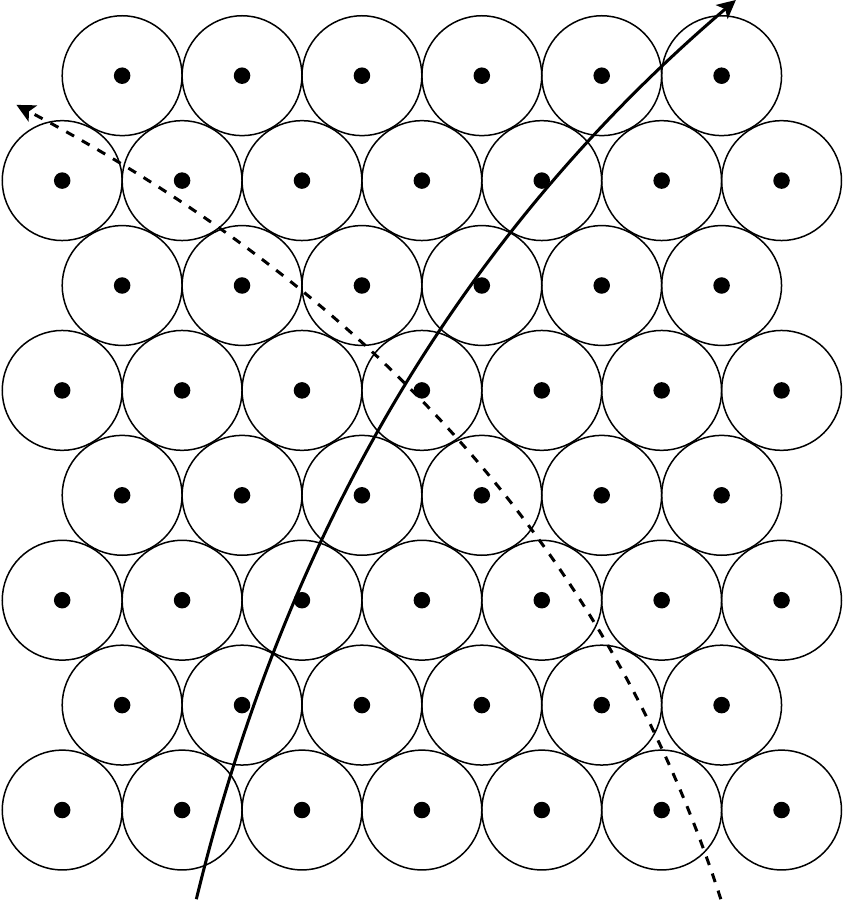}
        \caption{}
        \label{fig:ca1}
    \end{subfigure}\hfil
    \begin{subfigure}{0.2\textwidth}
        \includegraphics[width=\linewidth]{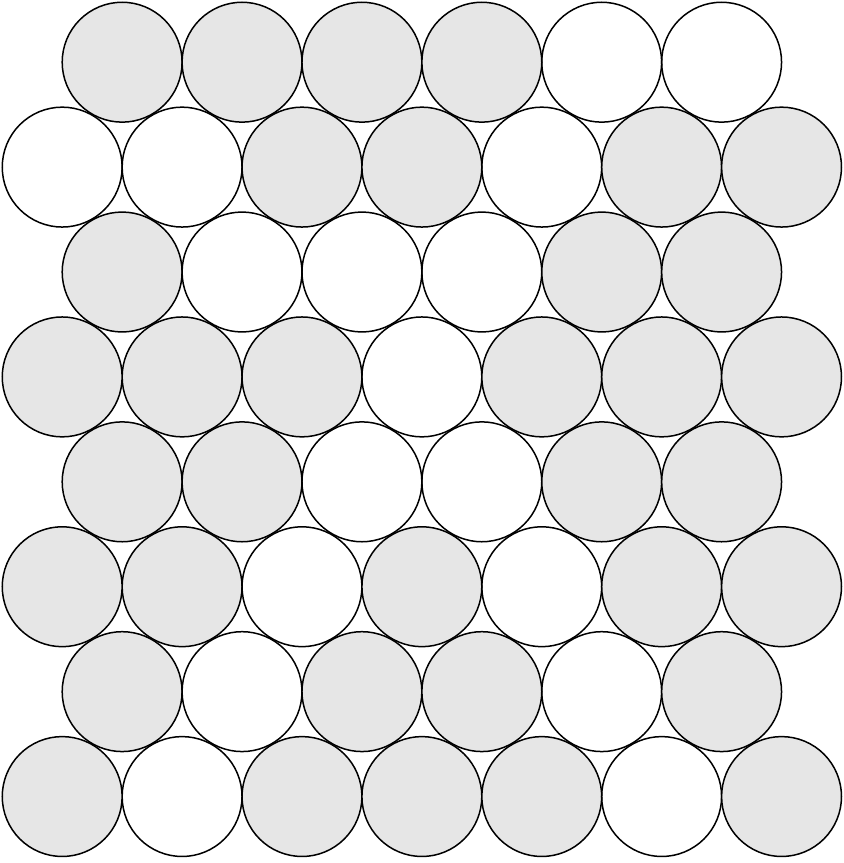}
        \caption{}
        \label{fig:ca2}
    \end{subfigure}\hfil
    % \medskip
    \begin{subfigure}{0.2\textwidth}
        \includegraphics[width=\linewidth]{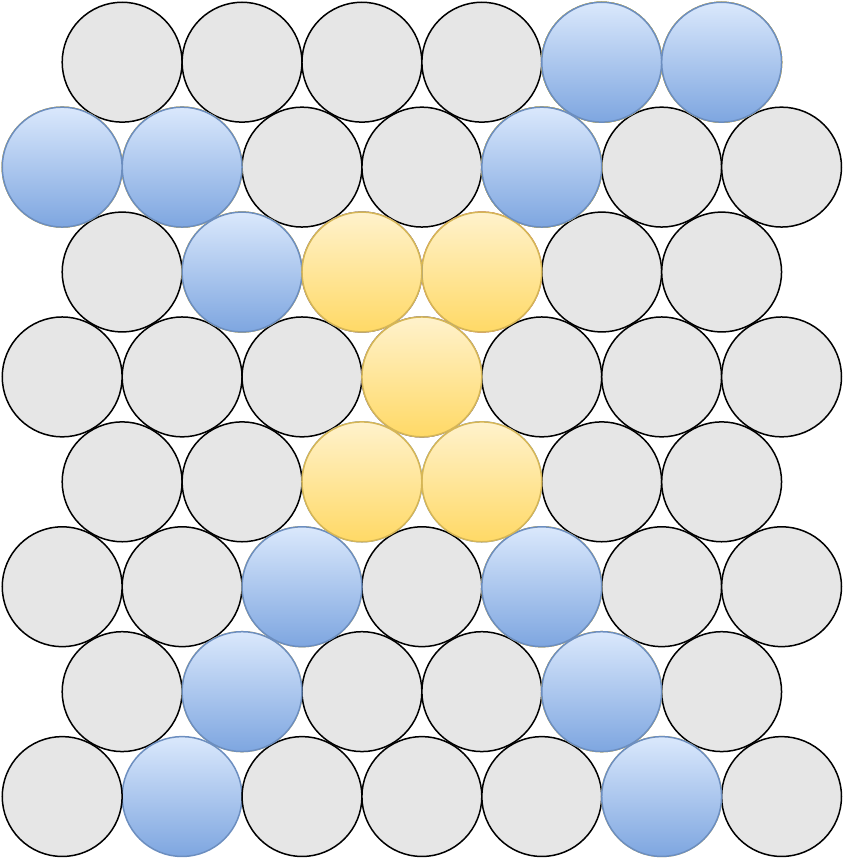}
        \caption{}
        \label{fig:ca3}
    \end{subfigure}\hfil
    \begin{subfigure}{0.2\textwidth}
        \includegraphics[width=\linewidth]{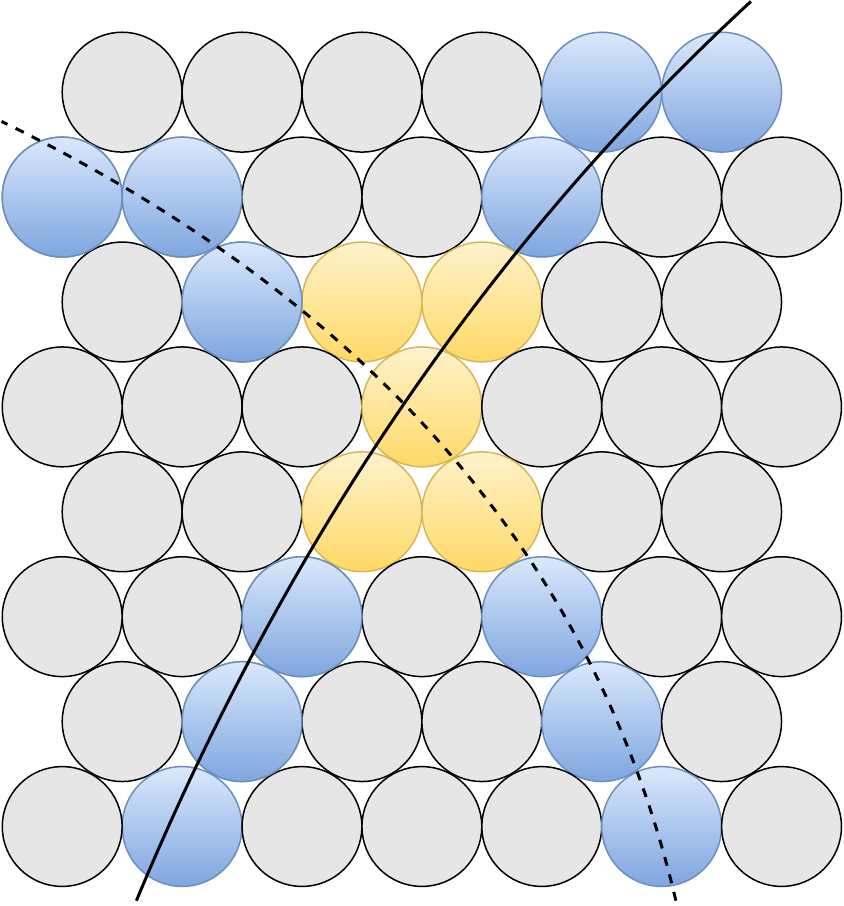}
        \caption{}
        \label{fig:ca4}
    \end{subfigure}
    \caption{Illustration of the Cellular Automaton workflow. (\subref{fig:ca1}) Particles traverse the detector. (\subref{fig:ca2}) Cells corresponding to detector hit are marked as active. (\subref{fig:ca3}) Active cells are classified as unambiguous (blue) or ambiguous (yellow). (\subref{fig:ca4}) Track fits are used to resolve ambiguities and merge tracklets.}
    \label{fig:cellular}
\end{figure*}

The procedure is illustrated in Fig.~\ref{fig:cellular}. After particles traverse the detector (Fig.~\ref{fig:cellular}(\subref{fig:ca1})), the cells that correspond to detector hits are activated (Fig.~\ref{fig:cellular}(\subref{fig:ca2})). All activated cells are categorized as \textit{unambiguous} (blue), if they have one or two neighboring activated cells, or \textit{ambiguous} (yellow), if they have more activated neighboring cells (Fig.~\ref{fig:cellular}(\subref{fig:ca3})). All unambiguous cells are assigned to one \textit{tracklet}, a preliminary track segment, and all ambiguous cells are assigned to the best matching track after a track fitting is performed (Fig.~\ref{fig:cellular}(\subref{fig:ca4})). The algorithm for selecting the neighborhood is shown in Alg.~\ref{alg:3d_ca}.
The reasoning behind the hit classifications is the following. If a hit is surrounded by no other hits, it is not considered to be part of a track or tracklet, but rather a random background hit. However, being surrounded by one or two other hits suggests that the hit is part of a track or tracklet that either ends in this hit or passes through. Any additional neighbours beyond two are usually caused by several tracks crossing each other or a random background hit occurring close to a track, hence the ambiguity. However, the classification is made independently of any cells outside of the neighborhood. In case of ambiguous hits within a track candidate, the algorithm produces separate track segments which are disconnected at the point of the ambiguous hit. Consequently, the algorithm creates many track segments that still have to be connected. Track segments that are connected via ambiguous hits can be connected using a preliminary parameter estimation of the tracks in the $xy$-plane by the Riemann fit. Based on these parameters, track segments are combined into common tracks.

%Introducing an extra dimension of time, the state of a cell is  updated step-wise in time according to the state of its neighboring cells.

\begin{algorithm}
    \KwData{STT hits}
    \KwResult{clustered STT hits}
    \Begin{
    \For{each STT hit}{
        \eIf{STT hit is neighbouring another STT hit}{
            accept hits as neighbours}{
            reject hits as neighbours}
    }
    \Return{clustered hits}
    }
    \caption{3D Cellular Automaton}
    \label{alg:3d_ca}
\end{algorithm}

%The algorithm needs to work for up to 10-15 concurrent tracks. This includes tracks from complex topologies such as tracks from secondary particles and low momentum particles curling in the magnetic field.

\subsection{Track reconstruction of free-streaming data}
\label{sec:FreeStreamingReco}

To adapt the reconstruction algorithms that work event-by-event to also work well for the free-streaming data with event-mixing, spatial information from the hits alone will not be sufficient to perform the pattern recognition (see Sec.~\ref{sec:free_streaming_data}). The timing information of the electronic signals has to be taken into account as well. 

In PandaRoot simulations, the time-stamp of a hit in the STT, $t_{hit}$, is calculated by the time at which a reaction occurred $t_{event}$, the travel time of a particle from its point of creation to the detector element that was hit $t_{travel}$, and the drift time of the electrons inside a straw tube from the point of closest approach between the particle trajectory and the tube's anode wire \textit{i.e.} the time it takes the first electrons to arrive, $t_{drift}$. The final expression becomes 
\begin{equation}
    t_{hit} = t_{event} + t_{travel} + t_{drift}.
    \label{eq:time_stamps}
\end{equation} 
Under realistic conditions in the PANDA experiment, the orders of magnitude for these values would be quite different. While $t_{event}$ would be continuously increasing, the difference between adjacent events varies because of the Poisson distributed time between two consecutive events with a high probability of very short time intervals between events (see Fig.~\ref{fig:time_between_events}). The travel time of a particle is considerably shorter with $\mathcal{O}(t_{travel}) = 0.01 - 1 \, \text{ns}$. It becomes negligible when compared to the drift time of the electrons, i.e. $\mathcal{O}(t_{drift}) = 100 \, \text{ns}$. However, with the mean time between events being of a similar or even smaller order of magnitude than the drift time, mixing of hit signal between two or more events becomes possible and even common. 

\begin{algorithm}
    \KwData{STT hits, threshold time}
    \KwResult{clustered STT hits}
    \Begin{
    \For{each STT hit}{
    \If{STT hit is neighbouring another STT hit}{
        \eIf{time stamps of both STT hits < threshold time}{
            accept hits as neighbours}{
            reject hits as neighbours}
    }}
    \Return{clustered hits}
    }
    \caption{4D Cellular Automaton}
    \label{alg:4d_ca}
\end{algorithm}

To perform tracking in this high occupancy environment, the approach in Sec.~\ref{sec:even_based_reconstruction} was modified to include the hit time stamps as given in Eq.~\ref{eq:time_stamps}. The resulting algorithm is shown in Alg.~\ref{alg:4d_ca}. If spatially neighboring hits occur closer in time than 250 ns, they are classified as neighbors, otherwise discarded and cannot be contained in the same tracklet.

\section{Quality assurance for track reconstruction on free-streaming data}

A class for evaluating the efficiency and other quality variables is available in PandaRoot in order to have a common basis for evaluating the performance of the PANDA tracking algorithms. The quality of the reconstruction algorithms is assessed by comparing the contents of the reconstructed tracks with their generated counterparts. It is not known, a priori, which generated track corresponds to the reconstructed track. The \textit{true} track is therefore defined as that to which the majority of the hits in the reconstructed track belong. 

A reference track set must be chosen in order to have relevant generated tracks to compare the reconstructed tracks to. This is done using an ideal track finder that uses the generated track with the hits to create a reconstructed track object from this. The performance is then only limited by the detector acceptance and efficiency. One benefit of creating such an object is that the hits in the ideal track and the realistically reconstructed track can be directly compared since it has the same data structure. Another benefit is that different hit conditions, that correspond to a minimum hit requirement in different detectors for a track to be reconstructed, can be imposed on the ideal track. This is needed since the efficiency should be normalized to those tracks that can be reconstructed in the relevant subdetectors. The condition used for this study is $\geq$ 6 STT hits.
 
The matching of the ideal track to the realistically reconstructed track is performed through a system of links that contain the information about where an object is stored and to which objects it relates to. For example, a hit contains links to the monte-carlo particles that created it. 

The hit finding efficiency is defined as 

\begin{equation}
    \epsilon_{hit} = \frac{N_{correct}}{N_{gen}},
\end{equation}

\noindent where $N_{gen}$ is the number of hits in the generated track and $N_{correct}$ is the number of found hits in the reconstructed track from the generated track. If $\epsilon_{hit}$ = 1, all hits from the generated track were found. The hit purity within a track is defined as 

\begin{equation}
    p_{hit}=\frac{N_{correct}}{N_{found}},
\end{equation}

\noindent where $N_{found}$ is the total number of found hits from any generated track and $N_{correct}$ is the number of found hits from the true generated track. If $p_{hit}$ = 1, only hits from the true generated track were found and if $p_{hit}$ $<$ 1, hits from additional generated tracks were found. If hits from generated tracks that are not the true track are included in the reconstructed track, these hits are referred to as \textit{impurities}. No noise hits were included in this study.

Track categories are defined in terms of the hit finding efficiency and purity:

\begin{enumerate}
    \item \textbf{Fully, purely, found}: all hits from a generated track were found and there are no impurities in the reconstructed track. $\epsilon_{hit}$ = 1 and $p_{hit}$ = 1.
    \item \textbf{Fully, impurely, found}: all hits from a generated track were found and at least 70$\%$ of all hits are from the true track. $\epsilon_{hit}$ = 1 and 0.7 $<$ $p_{hit}$ $<$ 1.
    \item \textbf{Partly, purely, found}: not all hits from a generated track were found and there are no impurities in the reconstructed track. 0.7 $<$ $\epsilon_{hit}$ $<$ 1 and $p_{hit}$ = 1.
    \item \textbf{Partly, impurely, found}: not all hits from a generated track were found and at least 70$\%$ of all hits are from the true track. 0.7 $<$ $\epsilon_{hit}$ $<$ 1 and 0.7 $<$ $p_{hit}$ $<$ 1.
    \item \textbf{Ghost}: no true generated track could be identified. $p_{hit}$ $<$ 0.7. 
    \item \textbf{Clone}: one generated track was reconstructed more than once. The track with the most hits from the true generated track is referred to as the real track and other reconstructed tracks which have been created from the same generated track are referred to as clones.
\end{enumerate}

\noindent The tracks in the first four categories constitute the set of correctly reconstructed tracks whereas the tracks in the final two categories are referred to as wrongly reconstructed or false tacks.

The track finding efficiency is defined as 

\begin{equation}
    \epsilon_{track} = \frac{N_{Reco}}{N_{Gen}},
\end{equation}

\noindent where $N_{Gen}$ is the number of generated tracks and $N_{Reco}$ is the number of reconstructed tracks. Ghost and clone tracks are not included in the calculation of the track finding efficiency. 

The two main differences between the event-sorted and time-sorted tracking quality assurance is 1) the way the data is read (see Sec.~\ref{sec:FreeStreamingData}) and 2) the way the matching between tracks is carried out. In the simulation for free-streaming data, the matching between the realistically reconstructed track and the ideal track is performed by matching a unique link to each track of the tracks to each other. In the event-by-event reconstruction, this is done using an index.

The benefit of the method for free-streaming data is that it is more general than the event-by-event method and works for both ways of retrieving the data. The disadvantage is that it is considerably slower using the full link system. However, since the quality assurance is run offline, this is not problematic. 

\section{Results}

\subsection{Data Samples}

In order to benchmark the algorithms, a data sample of 1000 hadronic background events based on the Fritiof Model (FTF) \cite{FTF1,FTF2} was generated with PandaRoot. This was done at four different interaction rates: 0.5, 1.0, 2.0 and 4.0 MHz. The intervals are chosen around 2.0 MHz to include the double interaction rate at 4.0 MHz and narrower intervals closer to 0.5 MHz where the behavior of the 4D algorithm is expected to be similar to that of the 3D algorithm. The interaction rates were chosen within an interval around 2.0 MHz which is the expected average interaction rate at the start of PANDA. The beam time structure and Poisson distributed interaction rate is included in the simulations. A beam momentum of 6.2 GeV/\textit{c} was chosen as a mid-range point to benchmark the algorithm. For a reference track set, the ideal track finder is used with the requirement that $\geq$ 6 STT hits are present in the track. Moreover, the data is sorted in time and then retrieved in bursts of 2000 ns for both the ideal and realistic tracking. The time interval mimics the beam revolution time.

There is no pre-filtering of tracks on \textit{e.g.} transverse momentum or point of origin in the simulations presented in this study. This means that challenging cases such as particles that are trapped in the magnetic field and curl as well as decay products from long-lived particles are included.

\begin{figure*}
    \centering
    \includegraphics[width=1.0\linewidth]{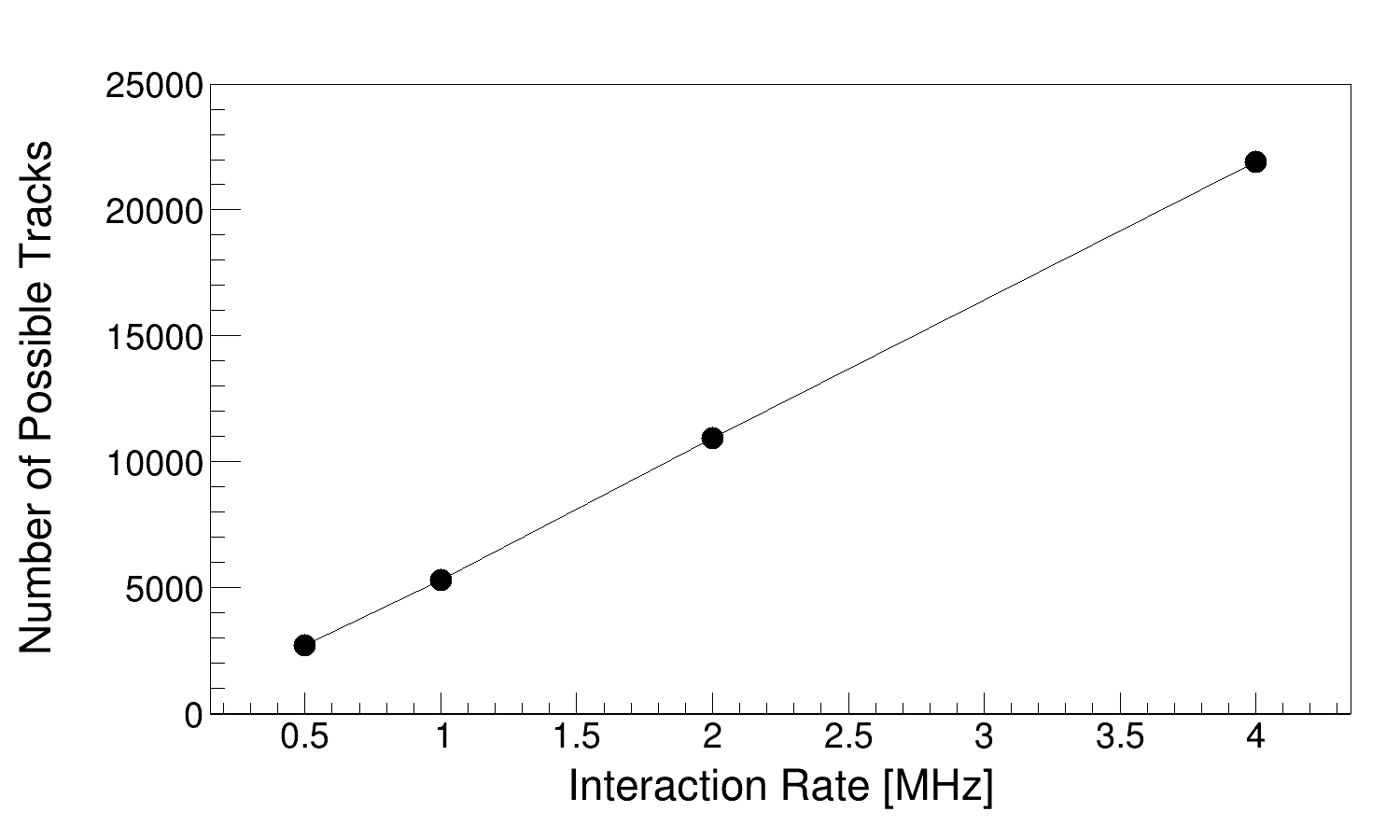}
    \caption{The number of reconstructed ideal tracks in 1000 bursts. Each burst contains data within a 2000 ns time interval.}
    \label{fig:num_tracks}
\end{figure*}

The number of reconstructed tracks within 1000 bursts at the different interaction rates can be seen in Fig. \ref{fig:num_tracks}. The number of ideal tracks scales linearly with the interaction rate. This is because of the increase in number of tracks contained within one burst when increasing the interaction rate allows for more data within the same interval due to the Poisson statistics. In order to reduce the runtime of the QA, 8557 tracks from 400 bursts are used in this data point to evaluate the efficiency of the reconstruction. This number is sufficient since the errors in the efficiencies are so small they are not visible in the graphs.

\subsection{Varying interaction rates}

\begin{figure*}
    \centering
    \includegraphics[width=0.49\linewidth]{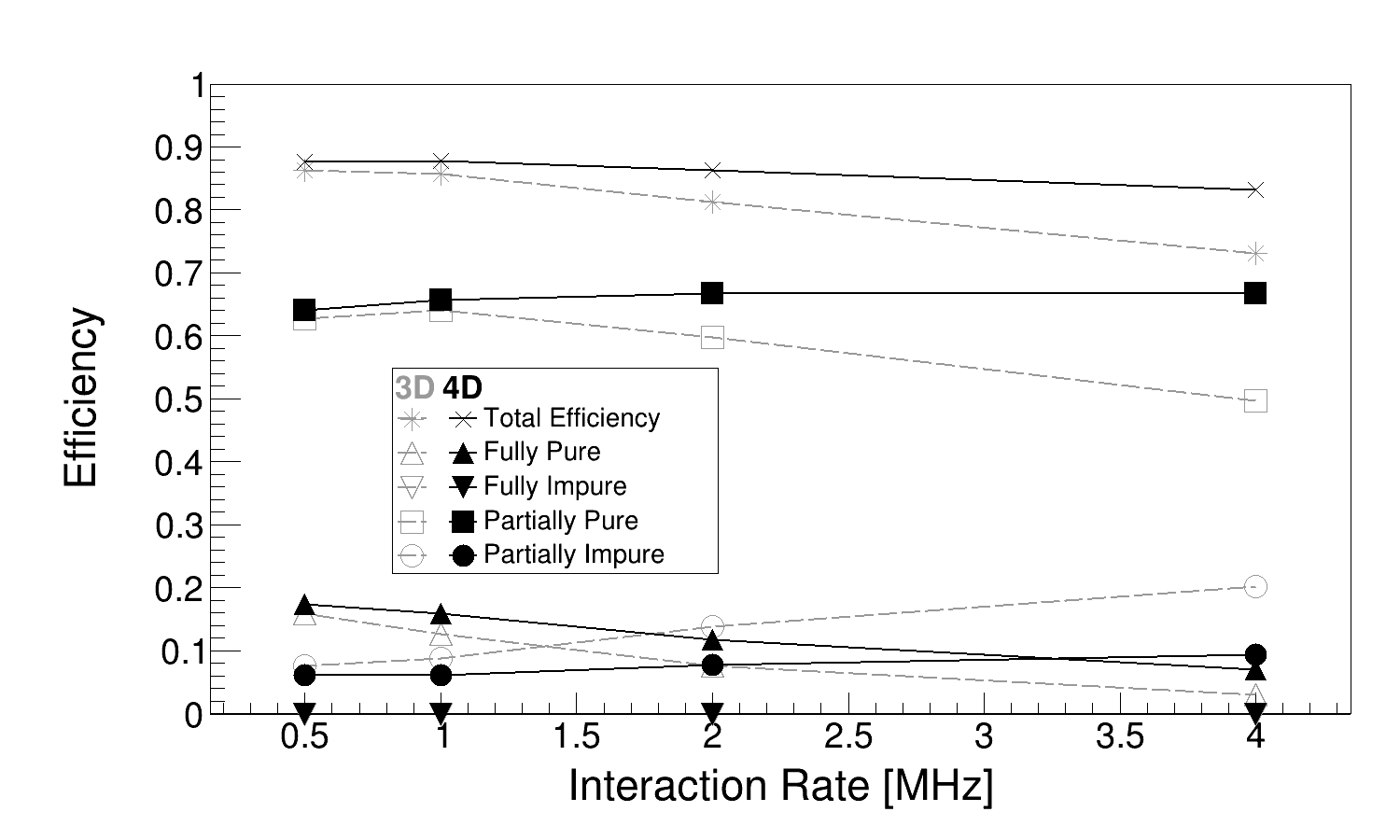}
    \includegraphics[width=0.49\linewidth]{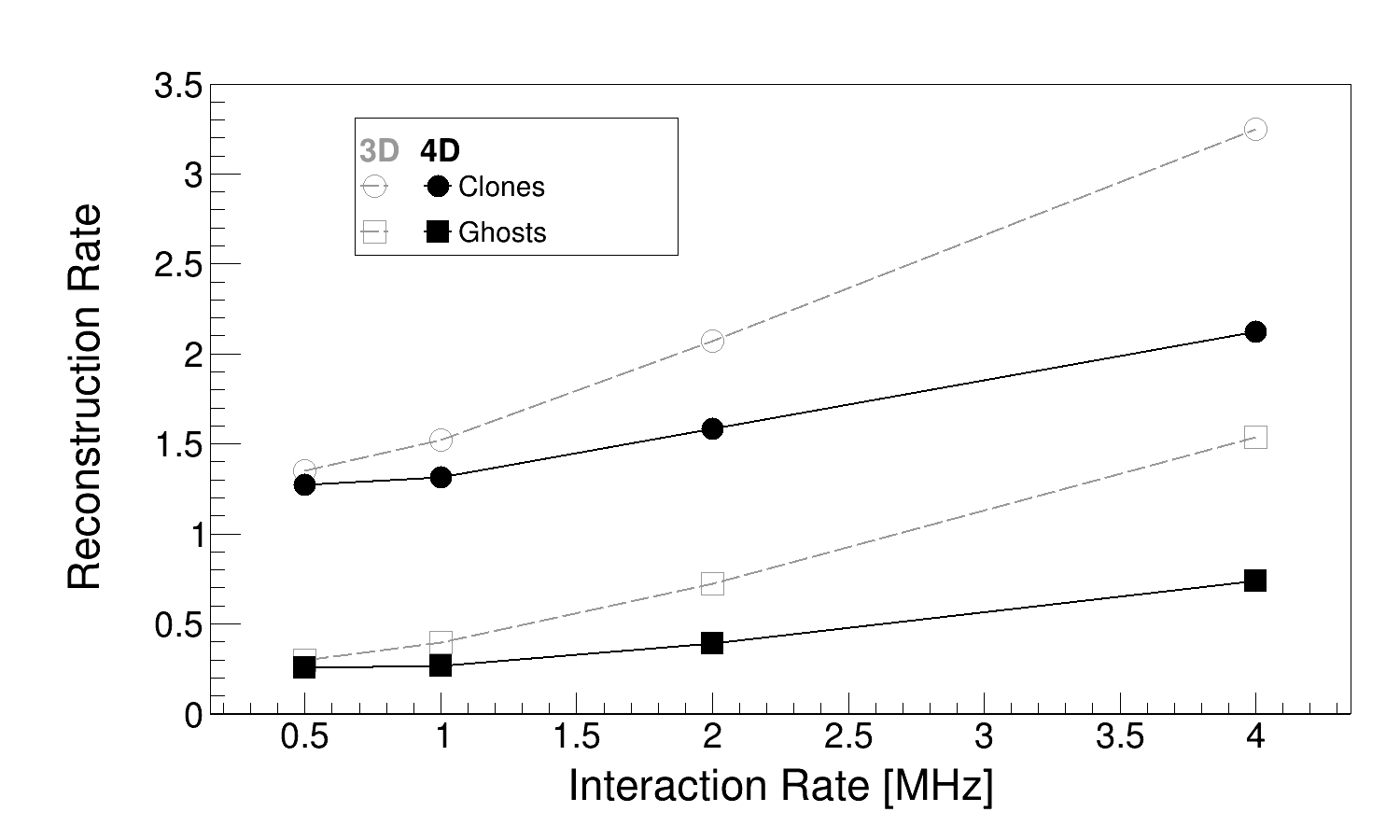}
    \caption{Efficiency (left) and the fake rate (right) for the 3D CA (dashed, gray, lines) and the 4D CA (solid, black, lines) vs the interaction rate. The time-cut for the 4D CA is 250 ns.}
    \label{fig:results_eff_fake_rates}
\end{figure*}

The efficiency is presented for both the 3D version of the CA as well as the 4D version for different interaction rates in the left panel of Fig.~\ref{fig:results_eff_fake_rates}. The efficiencies for the four track categories, fully pure, fully impure, partially pure and partially impure as well as the full efficiency are shown. At 0.5 MHz there is little temporal overlap between the events in the STT and the 3D and 4D version of the CA are expected to have a similar efficiency. It can be seen from the figure that this is also the case. However, at higher interaction rates the 4D version yields a larger total efficiency. The largest contribution to the full efficiency comes from partly pure tracks. The impure tracks contribute with almost 20$\%$ for the 3D CA at 4.0 MHz, but contributes with around 10$\%$ over the full range for the 4D CA. 

The fake rate (ghost and clone rate) is presented for varying interaction rates in the left panel of Fig.~\ref{fig:results_eff_fake_rates} for both the 3D and 4D version of the CA. The time-cut is 250 ns (see Sec.~\ref{sec:FreeStreamingReco}). The number of fake tracks in each category is normalised to the number of ideal true tracks. It is clear that including the time information decreases the fake rate and the effect is most prominent at higher interaction rates.

\subsection{Varying time-cut}

\begin{figure*}
    \centering
    \includegraphics[width=0.49\linewidth]{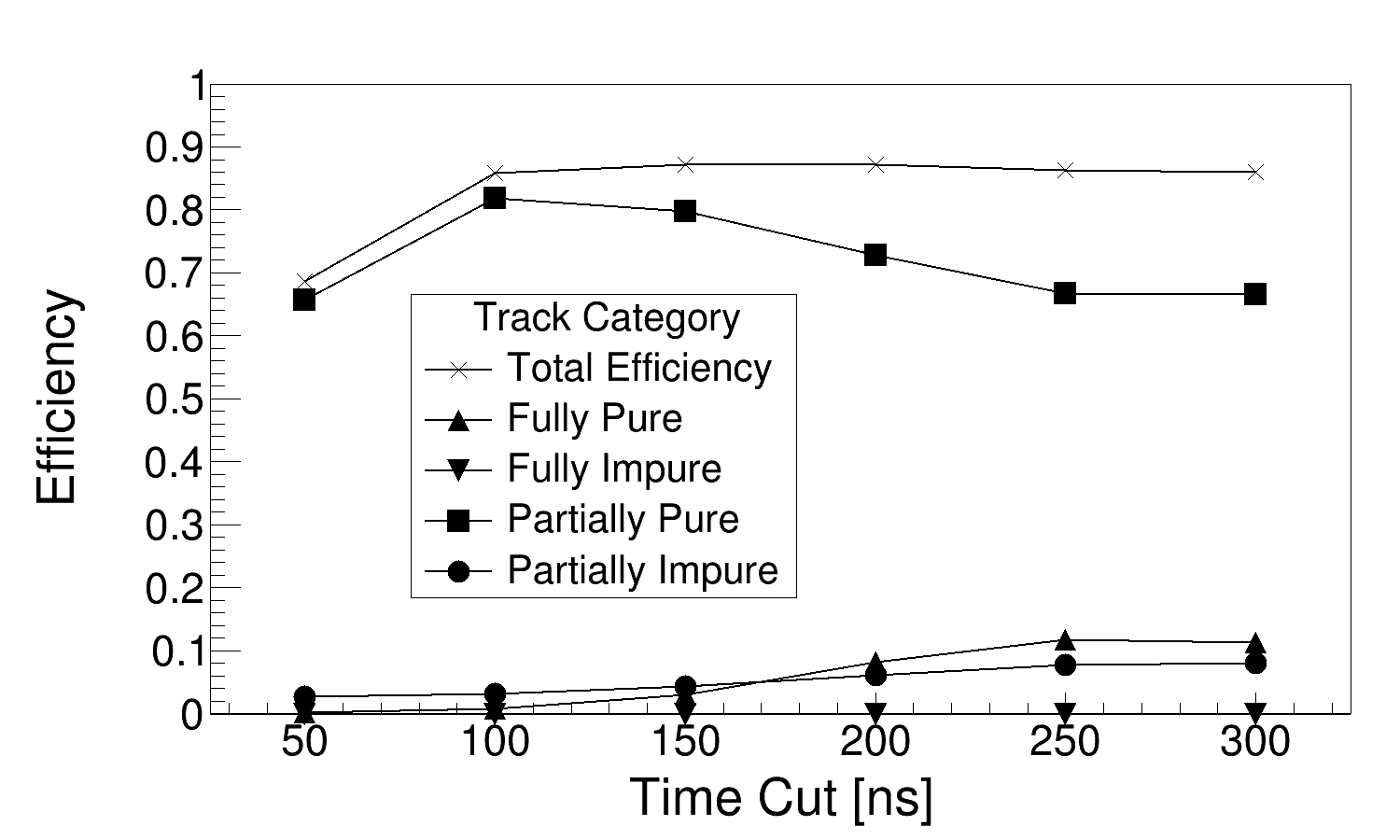}
    \includegraphics[width=0.49\linewidth]{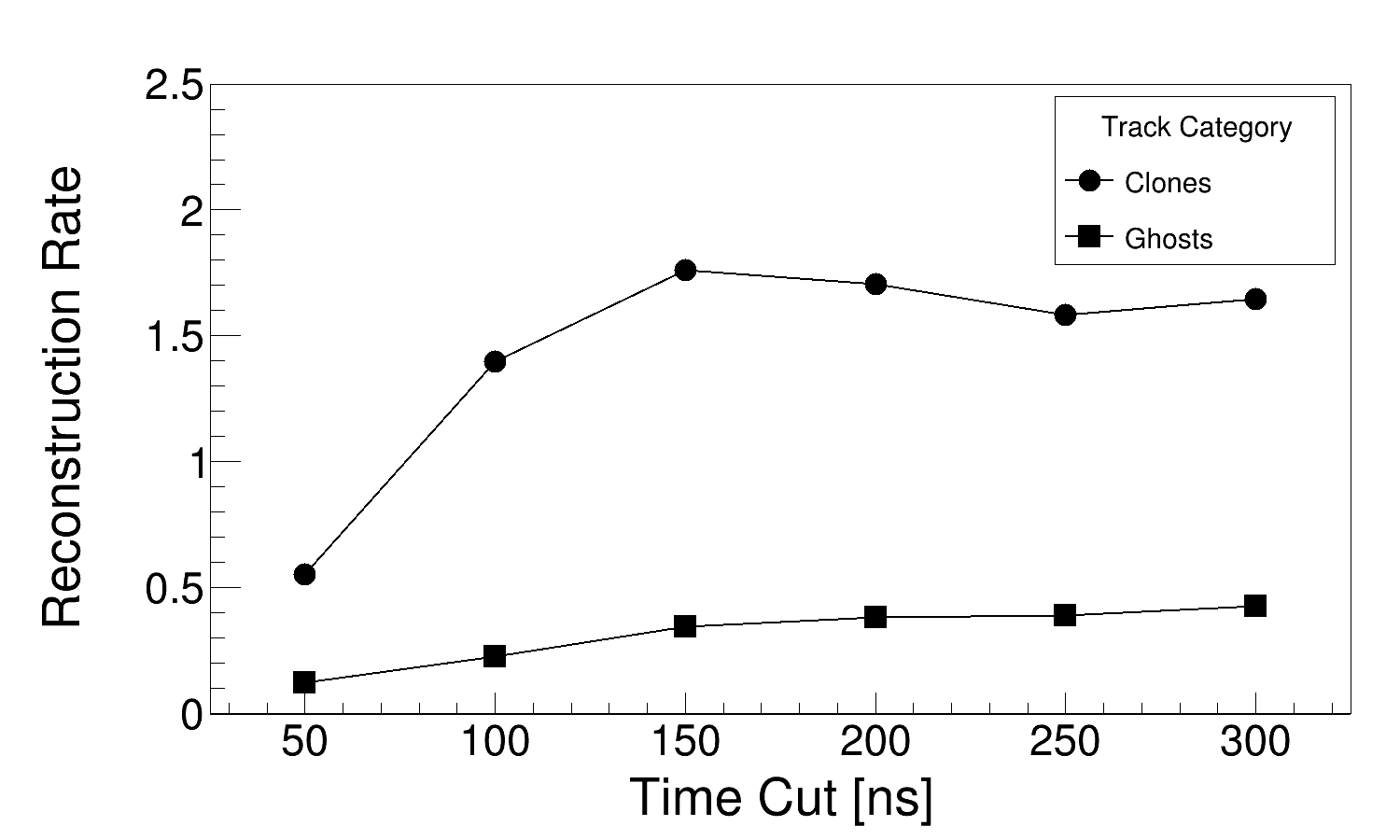}
    \caption{Efficiency (left) and fake rate (right) for the 4D CA vs the time-cut at 2.0 MHz interaction rate.}
    \label{fig:results_eff_timecut}
\end{figure*}

The effect of the time-cut value on the efficiency for the 4D CA is presented in Fig~\ref{fig:results_eff_timecut}, left panel. The interaction rate is 2.0 MHz in all data points. The efficiency is evaluated at a time-cut of the nominal 250 ns as well as 50, 100, 150, 200 and 300 ns. If the time-cut is below 250 ns, the number of fully pure tracks decrease while the number of partially pure tracks increase. Hence, the full efficiency remains stable. At a time-cut less than 100 ns, the full efficiency starts to drop since too few hits are allowed to form a tracklet. This is because the time-cut is so tight, true hits start to get discarded from the track.

The fake rate as a function of the value of the time-cut in the 4D CA at 2.0 MHz interaction rate can be seen in Fig.~\ref{fig:results_eff_timecut}, right panel. The fake rate, for both ghosts and clones, is reduced for stricter time-cuts. Therefore, a 250 ns time-cut is a good choice to keep the efficiency high while simultaneously suppressing the fake tracks. This value also corresponds to the maximum drift time of the electrons.

\subsection{Run-Time Performance}

\begin{figure}
    \centering
    \includegraphics[width=1\linewidth]{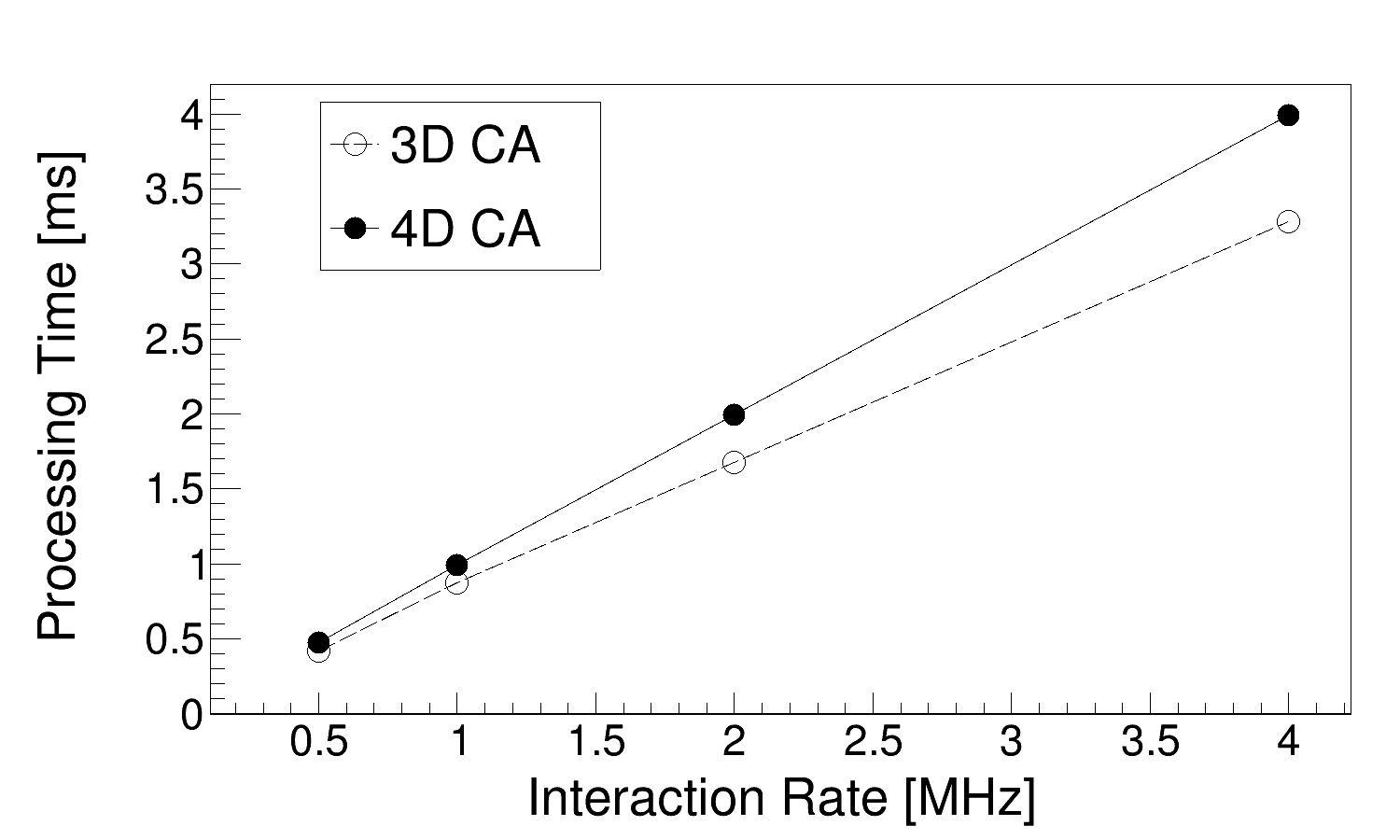}
    \caption{The time per event for running the full reconstruction for different interaction rates.}
    \label{fig:performance}
\end{figure}

Since the reconstruction algorithm is being developed to operate online, speed is a consideration. The computational performance for the full reconstruction is shown in Fig.~\ref{fig:performance} for the different interaction rates. The tests were performed on a laptop with the following hardware specifications:
\begin{itemize}
    \item Processor: Intel Core i7-1185G7, 3.00 GHz
    \item Memory: 32 GB RAM
\end{itemize} 

\noindent The time is calculated as the average processing time for 100 data bursts for each interaction rate. The increasing number of hits per burst makes the processing time per event longer at higher interaction rates. There is a linear dependence between the processing time and the interaction rate.  

The time-cut inclusion in Alg.~\ref{alg:4d_ca} is the only part that separates the 4D version from the 3D version. The run-time for this specific part of the code was therefore examined more closely. The increase in run-time when moving from the 3D version to the 4D is negligible compared to the full time for the reconstruction at all interaction rates. A larger increase occurs when going from the event-sorted reconstruction to the time-sorted and processing more data simultaneously in bunches.

\section{Conclusions and Outlook}

A Cellular Automaton-based track reconstruction algorithm and the tracking quality assurance procedure in PandaRoot has been adjusted to work for time-sorted data. The tracking algorithm has also been developed to work in 4D, taking into account both the spatial dimensions and time. A comparison between the 3D method and 4D method shows that at 4.0 MHz interaction rate, the efficiency is higher for the 4D version, 83$\%$, than for the 3D version with a 73$\%$ efficiency. In addition, the 4D method yields a much lower fake rate of tracks at higher interaction rates. 

Some improvements to the algorithm will be made in the future. The signal from a straw tube is characterized by two timestamps, the leading edge time, $t_{LE}$, when the signal in the readout electronics reaches above threshold and the trailing edge time, $t_{TE}$ when the signal drops below threshold. Currently, the procedure is used with $t_{LE}$. However, in the future it will be used with $t_{TE}$, when this is implemented in the software. This is because $t_{TE}$ is independent of the isochrone radius and therefore contributes with less uncertainty.

This is the first time the quality of a track reconstruction algorithm that works on free-streaming data for PANDA has been evaluated with the standard quality assurance of PandaRoot. Therefore, this marks an important step towards a realistic track reconstruction for PANDA. The quality assurance procedure will be useful for the developments of further algorithms for free-streaming data at PANDA.

\section{Conflict of Interest}
The authors declare that we have no conflicts of interest.

\section{Acknowledgments}
The authors would like to thank Peter Wintz and Gabriela P\'{e}rez Andrade for fruitful discussions about the STT readout.



\end{document}